\documentclass[prd,preprint,amsmath,amssymb,nofootinbib,eqsecnum]{revtex4}
\usepackage{latexsym,ifthen,graphics,color,epsfig,hyperref,mathrsfs}

\newcommand{\ie}{{i.e.}}

\newcommand{\eg}{{e.g.}}
\newcommand{\aka}{{a.k.a.}}

\newcommand{\viz}{{viz.}}
\newcommand{\wrt}{with respect to}

\newcommand{\rhs}{right-hand side}

\newcommand{\be}{\begin{equation}}
\newcommand{\ee}{\end{equation}}
\newcommand{\bea}{\begin{eqnarray}}
\newcommand{\eea}{\end{eqnarray}}
\newcommand{\beas}{\begin{eqnarray*}}
\newcommand{\eeas}{\end{eqnarray*}}

\newcommand{\bear}{\begin{array}{l}}
\newcommand{\eear}{\end{array}}

\newcommand{\bcf}{\begin{center}\begin{figure}}
\newcommand{\ecf}{\end{figure}\end{center}}

\newcommand{\bct}{\begin{center}\begin{table}}
\newcommand{\ect}{\end{table}\end{center}}

\newcommand{\ds}{\displaystyle}

\newcommand{\eq}[1]{(\ref{eq:#1})}
\newcommand{\eqn}[1]{equation~(\ref{eq:#1})}
\newcommand{\Eqn}[1]{Equation~(\ref{eq:#1})}

\newcommand{\eqs}[2]{(\ref{eq:#1}) and~(\ref{eq:#2})}

\newcommand{\sect}[1]{section~\ref{sec:#1}}
\newcommand{\Sect}[1]{Section~\ref{sec:#1}}
\newcommand{\sects}[2]{sections~\ref{sec:#1} and~\ref{sec:#2}}

\newcommand{\app}[1]{appendix~\ref{app:#1}}

\newcommand{\D}{d}

\newcommand{\Fint}[1]{\int \mathcal{D} #1 \,}

\newcommand{\der}[2]{\frac{d #1}{d #2}}

\newcommand{\fder}[2]{\frac{\delta #1}{\delta #2}}

\newcommand{\Or}{\mathrm{O}}

\newcommand{\order}[1]{\Or \bigl( #1 \bigr)}

\newcommand{\hf}{\frac{1}{2}}

\newcommand{\deltahat}[1]{
	\bar{\delta}(#1)
}

\newcommand{\SU}{\mathrm{SU}}

\newcommand{\pf}{\mathcal{Z}}

\newcommand{\cutoff}{K}

\newcommand{\ep}{C}
\newcommand{\dd}{\dot{\ep}}
\newcommand{\knl}[1]{\cdot {#1}\cdot}

\newcommand{\flow}{\Lambda \partial_\Lambda}
\newcommand{\op}{\hat{A}}
\newcommand{\Count}{\Delta}

\newcommand{\Sint}{S}
\newcommand{\Stot}{S^{\mathrm{tot}}}
\newcommand{\Stilde}{\tilde{S}}

\newcommand{\Svert}[1]{{S}_{#1}}

\newcommand{\Tact}{T}

\newcommand{\dual}{\mathcal{D}}

\newcommand{\dualshift}{\mathcal{E}}
\newcommand{\homog}{\mathcal{H}}
\newcommand{\homogv}[1]{\mathcal{H}_{#1}}

\newcommand{\coupled}{\mathcal{P}}
\newcommand{\fpop}{\mathscr{I}}

\newcommand{\classical}[3]{\fder{#1}{\phi} \knl{#2} \fder{#3}{\phi}} 
\newcommand{\quantum}[2]{\fder{}{\phi} \knl{#1} \fder{#2}{\phi}} 

\newcommand{\classicalvar}[3]{\fder{#1}{\varphi} \knl{#2} \fder{#3}{\varphi}} 
\newcommand{\quantumvar}[2]{\fder{}{\varphi} \knl{#1} \fder{#2}{\varphi}} 

\newcommand{\eop}{\mathcal{O}}
\newcommand{\marginal}{\eop_\mathrm{mar}}

\newcommand{\intconst}{B}
\newcommand{\const}{c}

\newcommand{\classifier}{\hat{\mathcal{M}}}

\newcommand{\hepth}[1]{hep-th/#1}
\newcommand{\hepph}[1]{hep-ph/#1}

\begin{document}

\title{An Extension of Pohlmeyer's Theorem}

\author{Oliver~J.~Rosten}

\affiliation{Department of Physics and Astronomy, University of Sussex, Brighton, BN1 9QH, U.K.}
\email{O.J.Rosten@Sussex.ac.uk}

\begin{abstract}
	Applying the Exact Renormalization Group to scalar field theory in Euclidean space of general
 	(not necessarily integer) dimension, 
	it is proven that the only fixed-point with vanishing anomalous dimension is the Gaussian 
	one. The proof requires positivity of the two-point connected correlation function together
	with a technical assumption concerning solutions of the flow equation.
	The method, in which the representation of the flow equation as a heat equation plays a 
	central role, extends directly to non-gauge theories with arbitrary matter content (though non-linear 
	sigma models are beyond the scope of the current method).
\end{abstract}

\maketitle

\section{Introduction}

Pohlmeyer's theorem~\cite{Pohlmeyer} gives a simple criterion for a massless field to be free. In particular, he showed that if a (real) scalar field, $\varphi$, lives in Minkowski space with any integer number of space dimensions then, assuming that the vacuum is unique, the theory is free if the two-point correlation function in canonical. The proof can be readily extended to non-gauge fields transforming under finite dimensional representations of the Lorentz group.

This result builds on earlier work by Jost~\cite{Jost} and, independently, Schroer and also Federbush and Johnson~\cite{F+J}. In 1960, Federbush \& Johnson proved, in a very simple way, that if the two-point correlation function of a massive scalar field agrees with that of a free field at equal times, then all correlation functions coincide with those of free fields. Slightly later, though independently, Jost (with reference to the unpublished work of Schroer) proved something rather similar (again in the massive case). First, it was shown that if the two-point correlation function agrees with that of a free field, then the field equation is that of a free field. Secondly, it was shown that if those $n$-point correlation functions with $n\leq 4$ agree with those of a free field then the commutation relation is that of a free field.
It is essentially the second point which distinguishes Jost's work from that of Federbush and Johnson, for Jost asserts that both of these criteria are necessary and sufficient to define a free field.

The language of these old proofs is that of axiomatic field theory. In this paper, we will take a completely different approach and, in the process, demonstrate something new. The formalism that we will employ is the Exact Renormalization Group (ERG), which developed from Wilson's groundbreaking insights into quantum field theory~\cite{Wilson}. The motivation for Wilson's work was to develop an understanding of systems exhibiting a large number of degrees of freedom per correlation length. He realized that if one could understand small patches of such a system then \emph{so long as the interactions are local} an understanding of the entire system can be built up by an iterated coarse-graining procedure (similar in spirit to the block-spinning---\aka\ blocking---of Kadanoff~\cite{Kadanoff}).
Consequently, it is reasonable to expect that a formalism based on this approach might be profitably applied to local quantum field theories.%
\footnote{Interestingly, a Wilsonian approach can be developed for non-commutative systems, 
where the non-locality can be sufficiently tamed by formulating everything in terms matrices~\cite{G+W-4D,RG+OJR}.
}

Along these lines, we can start by considering some quantum field theory defined at a bare scale, $\Lambda_0$. The bare action, $S_{\Lambda_0}$, respects any symmetries present and satisfies some locality constraints to be mentioned later but is otherwise arbitrary. To implement the continuum version of Kadanoff blocking, we start by dividing momentum modes into those above or below some effective scale, $\Lambda$. This requires that we work in Euclidean space. The next step is to integrate out degrees of freedom between the bare and effective scales. In the process, the bare action evolves into the Wilsonian effective action, $S_\Lambda$. The ERG (or flow) equation determines how $S_\Lambda$ changes with $\Lambda$.

Within the entire space of allowable actions, quantum field theorists pay particular attention to those which are renormalizable. In the Wilsonian approach, there are very simple conditions under which a theory is renormalizable \emph{nonperturbatively}. First, the theory can sit at a fixed-point of the ERG transformation, in which case the theory is independent of scale. Borrowing terminology from condensed matter physics, fixed-points supporting relevant directions are critical (and are generally the ones in which we are interested). Perturbing a fixed-point action in such a direction generates a flow along one  of Wilson's so-called renormalized trajectories. As the name suggests, such theories are renormalizable and, as noted by Morris~\cite{TRM-Elements}, it is easy to prove this nonperturbatively. 

Thus, fixed-points form the basis of renormalizable theories (we are ignoring the possibility of limit cycles or other exotic RG behaviour). Each critical fixed-point is characterized by an anomalous dimension, $\eta_\star$, (a $\star$ denotes fixed-point quantities) which encodes how the behaviour of the  two-point  connected correlation function differs from that which might be na\"{\i}vely inferred from the canonical dimension of the field, \viz
\[
	G(p) \sim \frac{1}{p^{2(1-\eta_\star/2)}}.
\]

In this paper we will prove---assuming  positivity of $G(p)$---that the only fixed-point with $\eta_\star = 0$ is the Gaussian one, for which the field is free. Unlike Pohlmeyer's theorem, this new proof takes place in Euclidean space, which is allowed to be of general (not necessarily integer) dimension, $\D$. It is worth noting that part of the purpose of this paper is to show that the Exact RG---which seems to be have been unfairly branded as inexact in some quarters due to a perceived necessity to immediately approximate the flow equation---can be used to prove things in quantum field theory.

The remainder of this paper is organized as follows. Aspects of the basic ERG formalism necessary for the extension of Pohlmeyer's theorem are provided in \sect{ERG}; to make this paper reasonably self contained, these elements are embedded in an overview of the ERG. It is in \sect{Linear} that   
a technical assumption required for the analysis is stated.
An extension of the formalism to
facilitate the computation of correlation functions---which is crucial for what follows---is given in \Sect{CFs}. Whilst the beginning of this section mimics that of~\cite{Fundamentals}, a refinement of the methodology is introduced (though the fine detail is relegated to \app{Alternative}). Combining insights from \sects{ERG}{CFs}, the extension of Pohlmeyer's theorem is quick and simple to prove, as described in \sect{Pohl}. To emphasise this simplicity, which is in danger of being obscured by the overhead in potentially unfamiliar ERG technology, the crucial steps of the argument are recapitulated in the conclusion, \sect{Conclusion}, after which possible future directions are indicated.

\section{The Exact Renormalization Group}
\label{sec:ERG}

\subsection{The Flow Equation}
\label{sec:flow}

The foremost ingredient of the exact renormalization group is a coarse-graining procedure which leaves the partition function unchanged~\cite{Wegner_CS,TRM+JL,mgierg1}. If the coarse-graining is carried by some $\Psi(p)$ (which must depend on the Wilsonian effective action~\cite{Wegner_CS,Fundamentals}) then  a large family of flow equations follows from
\be
-\flow e^{-\Stot_\Lambda[\varphi]} =  \int_p \fder{}{\varphi(p)} 
	\left\{
	\Psi(p) e^{-\Stot_\Lambda[\varphi]}
	\right\},
\label{eq:blocked}
\ee
where the reason for denoting the total action by $\Stot_\Lambda$ will become apparent in a moment.
The derivative \wrt\ $\Lambda$ is performed at constant $\varphi$, and the functional derivative
on the \rhs\ ensures that the partition function flows into the integral of a total derivative, which vanishes.
The only definite requirements on $\Psi$ are that it implements ultraviolet (UV) regularization and corresponds to a suitably local blocking procedure~\cite{aprop}.%
\footnote{Locality in this context---and throughout the paper (including the introduction)---should strictly be interpreted as quasi-locality. Applied to some function of momentum, say $f(p;\Lambda)$, this weaker condition is satisfied if $f$ has a Taylor expansion to all orders in $p^2/\Lambda^2$; thus non-analytic behaviour is forbidden, but functions such as $e^{-p^2/\Lambda^2}$ are admissible.}
As particularly emphasised by Morris and Latorre~\cite{TRM+JL}, flow equations following from~\eq{blocked} can be understood as originating from an infinitesimal field redefinition $\varphi \mapsto \varphi - \frac{\delta\Lambda}{\Lambda} \Psi$ for each RG step $\Lambda \mapsto \Lambda - \delta \Lambda$.

A particularly convenient flow equation was discovered by Polchinski~\cite{Pol}. This equation can be derived in many ways (see \eg~\cite{Pol,TRM-approxSolns,SalmhoferBook}); we will do so by first splitting the action in a particular way, which involves separating out what can generally be identified as a regularized kinetic term:
\be
	\Stot_\Lambda[\varphi] = \hf \varphi \cdot \ep^{-1} \cdot \varphi + \Sint_\Lambda[\varphi].
\label{eq:split}
\ee
Our notation is as follows: given a UV cutoff function, $\cutoff(p^2/\Lambda^2)$, [which decays faster than any power for $p^2/\Lambda^2 \rightarrow \infty$ and satisfies $\cutoff(0)=1$] we construct what is essentially a UV regularized propagator, $\ep(p^2;\Lambda^2) = \cutoff(p^2/\Lambda^2)/p^2$. The inverse of this object appears in $\varphi \cdot \ep^{-1} \cdot \varphi = \int_p \varphi(p)\ep^{-1}(p^2;\Lambda^2) \varphi(-p)$. Let us emphasise that the splitting~\eq{split} is just a matter of convenience: the interaction part of the action, $\Sint_\Lambda$, can quite legitimately contain two-point pieces which could even remove the $\order{p^2}$ part of what we have blithely called the kinetic term (though the resulting theory would, presumably, be non-unitary after continuation to Minkowski space~\cite{WeinbergI}).

A generalized version of Polchinski's equation can be derived by setting
\be
	\Psi(p) = \hf \dd(p^2;\Lambda^2) 
	\biggl[
		\fder{\Sint[\varphi]}{\varphi(-p)} - \ep^{-1}(p^2;\Lambda^2) \varphi(p)
	\biggr]
	+ \psi(p),
\label{eq:choice}
\ee
where $\dd(p^2;\Lambda^2) \equiv -\flow \ep(p^2;\Lambda^2)$ and $\psi$ gives us the freedom to perform additional field redefinitions along the flow. Note that we are no longer bothering to explicitly indicate the $\Lambda$-dependence of $\Sint$. It is now a simple matter to check, using~\eq{blocked}, that the choice~\eq{choice} yields, up to a neglected (divergent) constant term,
\be
	-\flow \Sint = \hf \classicalvar{\Sint}{\dd}{\Sint} - \hf \quantumvar{\dd}{\Sint} 
	+ \psi \cdot \fder{\Sint}{\varphi} - \fder{}{\varphi} \cdot \psi,
\label{eq:Pol-psi}
\ee
where \eg\ $\psi \cdot \delta \Sint /\delta \varphi = \int_p \psi(p) \delta \Sint /\delta \varphi(p)$. 
Throughout this paper, constant contributions to the action will be ignored.
Polchinski's equation is obtained simply by setting $\psi =0$. However, fixed-points are most conveniently uncovered by making a specific choice for $\psi$, as we will discuss in a moment.

The first step required to adapt the flow equation~\eq{Pol-psi} for the purpose of finding fixed-points is to transfer to dimensionless variables. Thus, we work with dimensionless coordinates, $\tilde{x} \equiv x\Lambda$, $\tilde{p} \equiv p/\Lambda$ and redefine the field by scaling out its canonical dimension: $\phi(\tilde{x}) = \varphi(x)/\Lambda^{(\D-2)/2} \,  \Rightarrow \, \phi(\tilde{p}) = \varphi(p)\Lambda^{(\D+2)/2}$. From a notational point of view, we will usually drop the tildes on the momenta: whether or not we are working in dimensionless variables will be implicit in whether we use $\phi$ or $\varphi$. With this in mind, the regularized two-point term, $\varphi \cdot \ep^{-1} \cdot \varphi$ becomes $\int_p \phi(p) \cutoff^{-1}(p^2) p^2 \phi(-p)$. We will denote the combination $\cutoff(p^2)/p^2$ by $\ep(p^2)$, or just $\ep$, mindful of the fact that this only follows from the dimensionful $\ep(p^2;\Lambda^2)$ after rescaling the momentum \emph{and} extracting a factor of $\Lambda^2$. So, as with $p$, what we mean by objects such as $\ep$ is to be determined by the context, this always being clear from whether it is $\phi$ or $\varphi$ which appears. When working in dimensionless variables, it is traditional to define the `RG
time', $t \equiv \ln \mu/\Lambda$, where $\mu$ is an arbitrary scale, with $-\flow = \partial_t$.

The rationale for rescaling to dimensionless variables is that the condition $\partial_t \Stot_\star = 0$ guarantees scale independence of the action~\cite{TRM-Elements,Fundamentals}. The structure of the rescaled flow equation is such that all couplings are dimensionless functions of $t$. Additional scales can creep in via a boundary condition for one or more of the couplings, say $g(t=t_0) = g_0$, but if we reside at a fixed-point, then the couplings are independent of scale and this possibility is precluded.

However, there is one more step necessary to adapt the flow equation so that it is suitable for finding all fixed-points. As recognized by Wegner~\cite{WegnerInv,Wegner_CS}, in particular (though see also~\cite{Wilson+Bell,Wilson+Bell-FiniteLattice,RGN}), the action contains both physical couplings and redundant couplings: the latter can be removed by appropriate quasi-local field redefinitions. Weinberg introduced the alternative nomenclature essential/inessential~\cite{Weinberg-AS} and noted that, because of their unphysical nature, there is no need for inessential couplings to stop flowing at what is, for the remaining couplings, a fixed-point (see also~\cite{Percacci-NewtonConstant,Percacci-RG+Units}). For example, the field strength renormalization, $Z$, can be removed from the action by introducing a reparametrization of the field along the flow, $\phi \mapsto \phi (1+\hf \eta \delta t)$ where, as usual, $\eta \equiv \Lambda d\ln Z/d \Lambda$.%
\footnote{Strictly speaking, we mean only to remove the field strength renormalization from the action up to a scale-independent constant. Therefore, in what follows, the full kinetic term need not be canonically normalized.}
This corresponds to choosing $\psi(p) = -\eta/2 \,\phi(p)$, yielding the flow equation
first written down by Ball et al.~\cite{Ball}:
\be
	\bigl(\partial_t -\hat{D}^- \bigr) \Sint
	= \classical{\Sint}{\cutoff'}{\Sint}
	-\quantum{\cutoff'}{\Sint}
	-\frac{\eta}{2} \phi \cdot \ep^{-1} \cdot \phi,
\label{eq:Ball}
\ee
where $\cutoff'(p^2) = d  \cutoff(p^2) /dp^2$ and, bearing in mind that throughout this paper will use a hat to denote a differential operator,
\be
	\hat{D}^{\pm} = \int_p
	\biggl[
		\biggl( \frac{\D+2 \pm \eta}{2} + p \cdot \partial_p \biggr) \phi(p)
	\biggr]
	\fder{}{\phi(p)}.
\label{eq:D_phi}
\ee

Of course, having removed $Z$ from the action, this begs the question as to why we do not remove other, or indeed all, redundant couplings from the action. The point is that the spectrum of $\eta_\star$, unlike the anomalous dimensions we would associate with the other redundant couplings, is quantized at critical fixed-points, meaning that such fixed-points with different values of $\eta_\star$ are indeed physically distinct. For anomalous dimensions whose spectrum is not quantized, we can go between apparently different fixed-points using infinitesimal, quasi-local field redefinitions. This point is discussed in greater detail in~\cite{Fundamentals} where, in addition, it is proven for the flow equation~\eq{Ball} that the spectrum of $\eta_\star$ at critical fixed-points is, indeed, quantized.

Let us conclude this section by noting that there are other ways of taking account of $Z$ in the flow equation, though it should be said that the method used above, advocated by H.~Osborn~\cite{HO-Remarks}, is particularly elegant. An alternative possibility is to take $\psi=0$, but shift $\Psi \rightarrow Z \Psi$; then after taking $\phi(\tilde{x}) = \varphi(x) /\Lambda^{(\D-2)/2} \sqrt{Z}$ we arrive at the same flow equation. The advantage of this latter method is that it makes is clear that we can essentially interpret the field, $\varphi$, as having a scaling dimension of $(\D-2+\eta)/2$. Indeed, when we henceforth talk of going from the dimensionless $\phi$ back to the dimensionful field, we understand this to include undoing the rescaling by $\sqrt{Z}$. Equivalently, we undo the accumulated field redefinition corresponding to $\psi$.

\subsection{Properties of Fixed-Points}

We will return to issues related to the quantization of $\eta_\star$ shortly. First, though, let us review some basic facts about fixed-points. The fixed-point criterion, $\partial_t \Sint_\star = 0$, implies that
\be
	\fpop(\Sint_\star) = 0,
\label{eq:fp-condition}
\ee
where
\be
	\fpop(\Sint_\star) = \classical{\Sint_\star}{\cutoff'}{\Sint_\star}
	-\quantum{\cutoff'}{\Sint_\star} + \hat{D}^-_\star \Sint_\star 
	-\frac{\eta_\star}{2} \phi \cdot \ep^{-1} \cdot \phi.
\label{eq:fp-operator}
\ee
In this paper we are interested in fixed-point solutions for arbitrary values of $\D$. We define what we mean by solutions for non-integer values of $\D$ as follows. To illustrate the basic idea, suppose for the moment that we can expand the action in powers of the field:
\[
	\Sint_\star[\phi] =
	\sum_n \int_{p_1,\ldots,p_n}
	\frac{1}{n!}
	\Svert{n}(p_1,\ldots,p_n;\Lambda) \phi(p_1)\cdots \phi(p_n)\deltahat{p_1+\cdots+p_n},
\]
where $\deltahat{p} \equiv (2\pi)^\D \delta^{\D}(p)$. At this level, we do not attempt to define
what we mean by $\D$-dimensional integrals over the fields. However, substituting this form
of the action into the fixed-point equation, we can strip off all fields to leave a tower of
coupled equations for the vertices, $\Svert{n}$, depending on $\D$.\footnote{This dependence on $\D$ comes from two sources: explicit dependence coming from $ \hat{D}^-_\star $ and also dependence coming from one-loop integrals generated by the second term on the \rhs\ of~\eq{fp-operator}. These one-loop integrals are of the form $\int_p \cutoff'(p^2) p^{2n}$, for integer $n$, and can be continued to arbitrary $\D$.} One can then search for solutions to this tower of equations for arbitrary values of $\D$.

The existence of a field-expansion of the action is not necessary. Indeed, instead of expanding in powers of the field, we could instead perform a derivative expansion (which is anyway more robust from the point of view of approximations). In this case, one finds a tower of coupled partial differential equations, again depending on the arbitrary parameter $\D$ (see for example~\cite{Ball}). Either way, we understand fixed-point solutions in non-integer dimensions to correspond to solutions of a tower of coupled equations, derived from~\eq{fp-condition}, which do not contain $\D$-dimensional integrals over the fields.

Having found some fixed-point, the RG eigenvalues, $\lambda_i$, together with the associated eigenoperators, 
$\eop_i$ are classified by perturbing the fixed-point action:
\be
	\Sint_t[\phi] = \Sint_\star[\phi] + \sum_i \alpha_i e^{\lambda_i t} \eop_i[\phi],
\ee
where the $\alpha_i$ are considered to be small.%
\footnote{In this context, `operators' are actually commuting functionals of the fields.}
Substituting this expression into the flow equation and working to first order in the $\alpha_i$ yields the eigenvalue equation
\be	
	\classifier \eop_i = \lambda_i \eop_i,
\ee
where
\be
	\classifier \equiv 2 \classical{\Sint_\star}{\cutoff'}{} - \quantum{\cutoff'}{} + \hat{D}^{-}_\star.
\label{eq:classifier}
\ee
Note that $\hat{D}^{-}_\star$ is obtained from~\eq{D_phi} by setting $\eta = \eta_\star$.
The relevant operators are those with $\lambda_i >0$, whereas the irrelevant ones have $\lambda_i <0$. Those with $\lambda_i = 0$ are marginal and one must go beyond linear order to determine whether or not they are exactly marginal or marginally relevant/irrelevant. Additionally, operators are classified according to whether or not they are redundant; as mentioned earlier, redundant operators correspond to infinitesimal field redefinitions and do not carry any physics.

As first recognized by Wegner~\cite{Wegner_CS}, if the spectrum of $\eta_\star$ is quantized, then the associated fixed-points each possess a marginal, redundant operator. For the flow equation we are using, this operator has been explicitly constructed by O'Dwyer and Osborn~\cite{HO+JOD}, though their expression is not required here. Rather, we will follow~\cite{Fundamentals} and notice that every fixed-point, critical or not, possesses a marginal operator given by
\be
	\marginal[\phi] = \hat{\Count} \Sint_\star[\phi], 
\ee
where
\be
	\hat{\Count} \equiv \hf \phi \cdot \fder{}{\phi} + \cutoff \cdot \fder{}{\cutoff}.
\label{eq:special}
\ee
That this is true is easy to see, upon recognizing that
\be
	\Bigl[\hat{\Count}, \classical{}{\cutoff'}{}\Bigr] = 0,
	\qquad
	\bigl[\hat{\Count}, \hat{D}^{\pm} \bigr] = 0.
\label{eq:SpecialCommutators}
\ee
With this in mind, it is apparent that
\begin{align}
	\classifier \, \hat{\Count} \Sint_\star & = 
	\hat{\Count}
	\biggl(
	\classical{\Sint_\star}{\cutoff'}{\Sint_\star} - \quantum{\cutoff'}{\Sint_\star} + 
	\hat{D}^{-}_\star \Sint_\star
	\biggr)
\nonumber
\\
	& = \hat{\Count} \biggl(\frac{\eta_\star}{2} \phi \cdot \ep^{-1} \cdot \phi \biggr)= 0.
\label{eq:emop}
\end{align}
The second line follows from the first after using the flow equation~\eq{Ball} specialized to a fixed-point.
At a critical fixed-point, the redundancy (or otherwise) of $\marginal$ was established in~\cite{Fundamentals}. For $\eta_\star <2 , \ \neq 0$, the operator is redundant; the same is true at the Gaussian fixed-point. Were other fixed-points to exist with $\eta_\star =0$, $\marginal$ would not, in general, be redundant; but of course this point is moot since we will show that such fixed-points do not exist. (From a physical perspective it is not surprising to see the constraint $\eta_\star<2$ arise, since it is required for critical fixed-points. In the context of the ERG, the reason for this was uncovered in~\cite{Fundamentals}.)

Irrespective of the redundancy of $\marginal$, this operator is exactly marginal~\cite{Fundamentals}: given some fixed-point, $\Sint_\star$, a line of related fixed-points can be generated according to
\be
	e^{b \hat{\Count}} \Sint_\star[\phi](\cdot) =  \Sint_\star[\phi](e^b),
\label{eq:line}
\ee
where $b$ is some real parameter, possibly with a restricted range. In the case that $\marginal$ is redundant, the fixed-points along this line are called equivalent, since neighbours are related to each other by an infinitesimal field redefinition. As it happens, we will exploit the techniques required to prove~\eq{line} to demonstrate the main result of \sect{Evaluation}. The details can be found in \app{Alternative}, where a proof of~\eq{line} is presented as a warmup exercise.

Let us illustrate some of the principles discussed above in the illuminating context of the Gaussian fixed-point. Clearly, there is a solution to~\eq{fp-condition} $\Sint_\star[\phi] = 0$, $\eta_\star=0$. However, it is easy to check that this is, in fact, only one representative of a line of equivalent fixed-points (obviously all with $\eta_\star =0$), which we will parametrize by $\intconst$:
\be
	\Sint_\star[\phi](B) = \hf \int_p \phi(-p) \frac{\intconst p^2}{1-\intconst \cutoff (p^2)} \phi(p)
	\quad \Rightarrow \quad
	\Stot_\star[\phi](B) =
	\hf
	\int_p \phi(-p) \frac{\ep^{-1}(p^2)}{1-\intconst \cutoff(p^2)} \phi(p).
\label{eq:Gaussian}
\ee
First, let us observe that we must take $\intconst <1$. This is because 
$\cutoff(0) = 1$ and we take $\cutoff(p^2)$ to be monotonically decreasing. So, were we to violate this bound, then the leading behaviour of the two-point vertex belonging to the total action would either cease to start at $\order{p^2}$ 
(for $\intconst = 1$) or would be of the wrong sign. Secondly, it is an easy matter to check that
\be
	\Sint_\star[\phi](B+\varepsilon)
	=
	\Sint_\star[\phi](B) + \frac{\varepsilon}{B} \hat{\Count} \Sint_\star[\phi](B) + \order{\varepsilon^2};
\label{eq:motion}
\ee
as anticipated, the marginal, redundant direction generates motion along the line of equivalent fixed-points.

By repeated application of $\hat{\Count}$ to $\Sint_\star$, it can be shown that
\be
	e^{b\hat{\Count}} \Sint_\star[\phi](\intconst) = \Sint_\star[\phi](\intconst e^b).
\label{eq:repeated}
\ee
Clearly, this is consistent with~\eq{motion}. Notice that if we start with a non-zero value of $\intconst$ (either positive or negative, but certainly less than unity) then the simplest representative of the Gaussian fixed-point, $\Sint_\star[\phi] = 0$, corresponds to $b \rightarrow -\infty$. This observation will be important later.

\subsection{Linearization}
\label{sec:Linear}

Further progress is achieved by recalling that 
the flow equation~\eq{Ball} can be linearized~\cite{Trivial}.%
\footnote{This is an exact statement, distinct from the notion of linearizing the flow equation in the vicinity of a fixed-point.} Were we to use the Polchinski equation written in dimensionful variables, it is easy to see that $e^{-\Sint_{\Lambda}[\varphi]}$ satisfies a heat equation. Rather than doing this, we will use the flow equation~\eq{Ball}. This equation can, in a sense, be solved  by introducing the `dual action',
\be
	\dual_t[\phi] = - \ln \bigl( e^{\op} e^{-\Sint_t[\phi]} \bigr),
\label{eq:Dual}
\ee
where, temporarily ignoring potential problems arising from infrared (IR) divergences,
\be
	\op \equiv \hf \classical{}{\ep}{}.
\ee
 As has been noted previously~\cite{Trivial,Fundamentals}, and as we will recall in \sect{CFs}, the dual action is intimately related to the correlation functions.

It is easy enough to check (see appendix~A of~\cite{Fundamentals}) that 
\be
	\left(
		\partial_t - \hat{D}^+ - \frac{\eta}{2} \phi \cdot \ep^{-1} \cdot \phi
	\right) e^{-\dual_t[\phi]} = 0.
\label{eq:DualFlow}
\ee
Let us note that although this equation is easy to solve, it is misleading to suppose that this makes the matter of finding quasi-local actions which solve~\eq{Ball} trivial: the majority of solutions to~\eq{DualFlow} are expected to correspond to actions which either violate quasi-locality or are ill-defined~\cite{Fundamentals}, or do not have a sensible limit as $t \rightarrow \infty$~\cite{HO-Remarks}. In some sense, we have traded the complicated equation \eq{Ball}, for which the boundary condition---a quasi-local action---is easy to implement, for the simple equation \eq{DualFlow}, for which the boundary condition for the action is very hard to implement (see~\cite{Fundamentals} for further discussion of this point).

With the basic structure now manifest, let us return to the issue of IR divergences. Should they be present, they arise due to the appearance of $\ep(p^2) = \cutoff(p^2) /p^2$
in $\op$. Indeed, in two dimensions, the Fourier transform of $\ep(p^2)$ blows up. Note, though, that this might not be as much of a problem as it sounds: if we supply IR regularization at intermediate steps then it might be that after $\op$ has acted on $\Sint$, sufficient powers of momenta are picked up to render the IR regularization unnecessary, in which case it can be removed.
With this in mind, let us introduce an IR scale, $k$, and an associated dimensionless parameter analogous to $t$: $s \equiv \ln \mu/k$. Utilizing this new scale, we define
\be
	\cutoff_{s-t}(p^2) \equiv \cutoff \bigl(p^2 e^{2(s-t)}\bigr),
	\qquad
	\ep_{s-t}(p^2) \equiv \frac{\cutoff(p^2) - \cutoff_{s-t}(p^2)}{p^2},
	\qquad
	\op_{s-t} \equiv \hf \classical{}{\ep_{s-t}}{}.
\ee
Since $\cutoff(0) = 1$, we see that the potentially dangerous $1/p^2$ is rendered harmless. 
This leads us to define
\be
	\dual_{t,s}[\phi] \equiv - \ln \bigl( e^{\op_{s-t}} e^{-\Sint_t[\phi]} \bigr),
\label{eq:Dual_k}
\ee
and so~\eq{DualFlow} becomes:
\be
	\biggl(
		\partial_t - \hat{D}^+ - \frac{\eta}{2} \phi \cdot \ep^{-1} \cdot \phi
		+ \eta \phi \cdot \frac{\cutoff_{s-t}}{\cutoff} \cdot \fder{}{\phi}
		+\frac{\eta}{2} \classical{}{\frac{\cutoff_{s-t} \ep_{s-t}}{\cutoff}}{}
	\biggr) e^{-\dual_{t,s}[\phi]} = 0.
\label{eq:DualFlow_k}
\ee

For all that follows, we will assume that $\lim_{s \rightarrow \infty} \dual_{t,s}[\phi]$ exists. However, even given this assumption, it remains to be shown that we can identify $\dual_{t,\infty}[\phi]$ with $\dual_t[\phi]$, which we now understand to be given by solutions of~\eq{DualFlow} rather than~\eq{Dual}, per se. Happily, in the case of $\eta =0$---which is of primary interest in this paper---this identification follows directly from our assumption. Indeed, setting $\eta = 0$ in~\eq{DualFlow_k}, we see that the equation for $\dual_{t,s}[\phi]$ is the same as the one for $\dual_t[\phi]$ [\ie~\eq{DualFlow} with $\eta = 0$]. Consequently, a solution $\dual_{t,s}[\phi]$ which survives the $s \rightarrow \infty$ limit must be a solution of~\eq{DualFlow} (with $\eta=0$). Therefore, in this case, we identify $\lim_{s \rightarrow \infty} \dual_{t,s}[\phi] = \dual_t[\phi]$.

Treating $\eta \neq 0$ is not so important for this paper, but is nevertheless desirable since it will be useful to compare certain fixed-point results coming from $\eta_\star =0$ to those with $\eta_\star \neq 0$. To this end, let us return to~\eq{DualFlow_k} and consider taking $s$ to be large.
In this regime, $\cutoff_{s-t}(p^2)$ dies off rapidly and so our na\"{\i}ve expectation might be that the final two terms in the large brackets are sub-leading. This is certainly true of the penultimate one. However, in the final term,  $\lim_{s \rightarrow \infty} \ep_{s-t}(p^2)$ can generate IR divergences which compensate the vanishing of $\cutoff_{s-t}(p^2)$.%
\footnote{%
This can be illustrated by considering the following one-dimensional integral designed to mimic $\int_p  \cutoff_{s-t}(p^2) \ep_{s-t} (p^2) / \cutoff(p^2)$:
$
I(a,\epsilon) = \int_\epsilon^\infty dz e^{-z/a+z} \bigl(e^{-z} - e^{-z/a} \bigr) /z ,
$
where we have chosen an exponential UV cutoff and have introduced an IR cutoff, $a$. The reason that we have set the lower limit to be $\epsilon$ is so that we can evaluate the integral in terms of $E_1(y) = \int_y^\infty e^{-z} /z \, dz = -\gamma -\ln y + \order{y}$, where $\gamma$ is the Euler-Mascheroni constant. Combining terms, it is easy to show that $I(a,0) = \ln (2-a)$, which manifestly does not vanish in the limit $a\rightarrow 0$.} 
So, if we are to be able to identify $\dual_{t,\infty}[\phi]$ with a solution of~\eq{DualFlow}, it must be that $\lim_{s\rightarrow \infty} \delta/\delta\phi \cdot \cutoff_{s-t} \ep_{s-t} / \cutoff \cdot \delta \dual_{t}[\phi] /\delta \phi  = 0$. We will henceforth assume this to be true, and will shortly see what this implies when we search for fixed-point solutions.

Let us begin our analysis of fixed-points by recognizing that
\[
	\partial_t \Sint_\star = 0
	\qquad
	\Rightarrow
	\qquad
	\partial_t \dual_{t,s}=
	-e^{2(s-t)} 
	e^{\dual_{t,s}} \, \fder{}{\phi} \cdot  \cutoff'_{s-t} \cdot \fder{}{\phi} \, e^{-\dual_{t,s}}.
\]
Recalling that the cutoff function dies off faster than any power of $e^s$---and noting, crucially, that $\cutoff'(p^2)$ is IR safe without any need for regularization---it is clear (given our assumption that $\dual_{t,\infty}$ exists) that
 $\partial_t \Sint_\star = 0 \ \Rightarrow \ \partial_t \dual_{\star,\infty}=0$. Consistent with the assumptions described above, we identify $\dual_{\star,\infty}= \dual_\star$, where the latter is understood as a solution of
 \be
	\left(
		 \hat{D}^+_\star + \frac{\eta_\star}{2} \phi \cdot \ep^{-1} \cdot \phi
	\right) e^{-\dual_\star[\phi]} = 0.
\label{eq:fp-dual}
\ee
To proceed, let us multiply on the left by $e^{\hf \phi \cdot h \cdot \phi}$, where $h(p^2)$ is to be determined. Indeed, if we choose $h$ to satisfy
\be
	-\frac{2+\eta_\star}{2} h(p^2) + p^2 h'(p^2)  + \frac{\eta_\star}{2} \ep^{-1}(p^2) = 0,
\label{eq:h-eq}
\ee
(where, as before, a prime denotes a derivative) then, defining 
\be
	\homog[\phi] \equiv  - \hf \phi \cdot h \cdot \phi + \dual_\star[\phi],
\label{eq:homog}
\ee
it is easy to check that
\be
	\hat{D}^{+}_\star \,  e^{-\homog[\phi]} = 0,
	\qquad
	\Rightarrow
	\qquad
	\hat{D}^{+}_\star \,  \homog[\phi] = 0.
\ee
The solutions to this equation are simple: $\homog[\phi]$ has an expansion in powers of the field with the vertices transforming homogeneously (hence $\homog$) with momenta:
\be
	\homog[\phi] =
	\sum_n \frac{1}{n!} \int_{p_1,\ldots,p_n} \homogv{n}(p_1,\ldots,p_n) \, \phi(p_1) \cdots \phi(p_n)
	\deltahat{p_1+\cdots+p_n},
\label{eq:Homog-solution}
\ee
where, for a scaling parameter $\xi$,
\be
	\homogv{n}(\xi p_1,\ldots, \xi p_n) = \xi^r \homogv{n}(p_1,\ldots, p_n),
	\qquad
	r = \D - n \frac{\D - 2 - \eta_\star}{2}.
\label{eq:r}
\ee
All that remains, then, to find $\dual_\star$ is to solve~\eq{h-eq}:
\be
	h(p^2) =
	-\tilde{\const}_{\eta_\star} p^{2(1+\eta_\star/2)}
	- \frac{\eta_\star}{2} p^{2(1+\eta_\star/2)}
	 \int^{p^2} dq^2 \frac{\cutoff^{-1}(q^2)}{q^{2(1+\eta_\star/2)}},
\label{eq:h-solution}
\ee
where $\tilde{\const}_{\eta_\star}$ is an integration constant, one for each value of $\eta_\star$. The integration constant is chosen as follows. First let us note that, for some other constants $\intconst_{\eta_\star}$,
\be
	\homog_2(p) \equiv \homog_2(p,-p) = -\intconst_{\eta_\star}  p^{2(1+\eta_\star/2)}
\label{eq:constants}
\ee
(the reason for the choice of sign will become apparent later).
We choose the $\tilde{\const}_{\eta_\star}$ by demanding that $h(p^2)$ has no pieces exhibiting this momentum dependence. For the special case of $\eta_\star = 0$ this implies that
$ h(p^2) = 0$, from which it follows that $\tilde{\const}_0 = 0$.%
\footnote{Strictly speaking, we should allow for the possibility that there are distinct fixed-points with the same value of $\eta_\star$. We can easily allow for this by furnishing $\intconst_{\eta_\star}$ with an extra label, though will not bother to indicate this explicitly. Notice that there is no need to do this for the $\tilde{\const}_{\eta_\star}$.}

It will prove useful to recast~\eq{h-solution} by integrating by parts:
\be
	h(p^2) =
	-\tilde{\const}_{\eta_\star} p^{2(1+\eta_\star/2)}
	+
	\ep^{-1}(p^2)
	-
	p^{2(1+\eta_\star/2)} 
	\int^{p^2} dq^2
	\left[
		\frac{1}{\cutoff (q^2)}
	\right]'
	q^{-2(\eta_\star/2)}.
\label{eq:h}
\ee
Now, so long as $\eta_\star <2$, we can define
\be
	\rho(p^2) \equiv 
	\ep^{-1}(p^2)
	- p^{2(1+ \eta_\star/2)}  \int_0^{p^2} dq^2
	\left[
		\frac{1}{\cutoff (q^2)}
	\right]'
	q^{-2(\eta_\star/2)}.
\label{eq:rho}
\ee
At the level of this equation, the restriction to $\eta_\star<2$ is necessary to ensure that the integral does not blow up at its lower limit. But, as mentioned earlier, it is not surprising to see such a constraint arise.
Using~\eq{rho} we rewrite~\eq{h} as follows:
\be
	h(p^2) = -\const_{\eta_\star} p^{2(1+\eta_\star/2)}
	+ \rho(p^2),
	\qquad
	\eta_\star <2,
\label{eq:h-recast}
\ee
where the $\const_{\eta_\star}$ are constants are related to the $\tilde{\const}_{\eta_\star}$. 
As we will confirm in a moment, it is easy to see that
\be
	\const_{\eta_\star} = 
	\left\{
		\begin{array}{ll}
			1, \ & \eta_\star = 0
		\\
			0, \ & \eta_\star <2, \ \neq 0.
		\end{array}
	\right.
\label{eq:c}
\ee
The first case is simple to check upon recognizing that, for $\eta_\star = 0$, $\rho(p^2) = p^2$. The second case follows upon exploiting the quasi-locality of the cutoff function:
\be
	\rho(p^2) = p^2 + \order{p^4},
\label{eq:rho-Taylor}
\ee
making it immediately apparent that, for $\eta_\star <2, \ \neq 0$, this term cannot supplement the $\const_{\eta_\star}$ piece.

Before moving on, let us consider the IR behaviour of $\dual_\star$. For $\eta_\star \geq 0$, the two-point contribution to the dual action is IR safe. For $\eta_\star <0$, we require $d+ 2 + \eta_\star >0$
in order that $\int_p \phi(p) \phi(-p) p^{2(1+\eta_\star/2)}$ is well behaved for small momentum.
Beyond the two-point level, we see from~\eqs{Homog-solution}{r} that, so long as no one momentum comes with too great an inverse power, $\dual_\star[\phi]$ as a whole is IR finite. 

We are now in a position to argue that, given certain restrictions, $\lim_{s \rightarrow \infty} \dual_{t,s}$ evaluated at a fixed-point coincides with $\dual_\star$ even for $\eta_\star \neq 0$. To this end, let us try to find solutions to~\eq{DualFlow_k} of the form $\dual_{t,s} = \dual_\star + \ldots$. We know that the contribution created when the penultimate term acts on $\dual_\star$ vanishes in the limit $s \rightarrow \infty$. Treating the final term is more subtle. Using~\eq{Homog-solution}, it is apparent that allowing the final term to act on $\dual_\star$ can produce things like
\[
	\int_q
	\frac{\cutoff_{s-t}(q^2) \ep_{s-t}(q^2)}{\cutoff(q^2)}
	\homog_n(q,-q,\ldots).
\]
Let us suppose that there is a contribution to $\homog_n(q,-q,\ldots) $ which is independent of $q$ [this is perfectly consistent with \eq{r}]. Then, if $\D>2$, the integral has no need for any IR regularization and so the above term vanishes if we take $s\rightarrow \infty$. Of course, there will be contributions to $\homog_n(q,-q,\ldots)$---which is symmetric under permutations of its arguments---that do depend on $q$. But, so long 
as individual momenta do not come with too big a negative power, then IR finiteness is maintained in the absence of $k$ and there are no contributions which survive the limit $s \rightarrow \infty$. Similar considerations apply in $\D \leq 2$, but now it is necessary that $\homog_n(q,-q,\ldots) $ does not have any pieces independent of $q$ [which is also perfectly consistent with~\eq{r}] in order that the IR regularization can be removed.
For what follows, we tacitly assume that $\dual_\star[\phi]$ is such that the IR behaviour of its vertices is sufficiently good.  This guarantees that, in the limit $s \rightarrow \infty$, \eqn{DualFlow_k} is solved by $\dual_\star[\phi]$.

Let us now consider the effect induced on the dual action by operating on $\Sint_\star$ with $e^{b\hat{\Count}}$. To do this, we use the result [as can be checked by Taylor expansion and using~\eq{special}] that, for some functional $X[\phi]$,
\be
	\exp \Bigl( -e^{b\hat{\Count}} X[\phi] \Bigr)
	= e^{b\hat{\Count}} e^{-X[\phi]}.
\ee
Employing this, together with the result $\bigl[\hat{\Count},\op \bigr] = 0$ (or $\bigl[\hat{\Count},\op_{s-t} \bigr] = 0$, if one prefers), it follows almost immediately
that
\be
	\Sint_\star[\phi] \mapsto e^{b\hat{\Count}} \Sint_\star[\phi] 
	\qquad
	\Rightarrow
	\qquad
	\dual_\star[\phi]  \mapsto e^{b\hat{\Count}} \dual_\star[\phi] .
\label{eq:line-dual}
\ee
We now proceed to evaluate the effect of $e^{b\hat{\Count}}$ on $\dual_\star$. First, we note that since the vertices of $\homog$ are homogeneous in momenta, they cannot depend on the cutoff function. Therefore, using~\eq{special}, we have that
\begin{align}
	e^{b \hat{\Count}} \homog[\phi](\cdot) &=
	\sum_n \frac{1}{n!} e^{bn/2}
	\int_{p_1,\ldots,p_n} \homogv{n}(p_1,\ldots,p_n) \, \phi(p_1) \cdots \phi(p_n)
	\deltahat{p_1+\cdots+p_n}
\nonumber
\\
	& =
	\homog[\phi](e^b) = \homog[e^{b/2} \phi ],
\label{eq:homog-operated}
\end{align}
where we recognize from the final expression that the effect of $e^{b \hat{\Count}}$ on $\homog[\phi]$ is to rescale the field. It is trivial to see that
\be
\lim_{b \rightarrow -\infty}\homog[\phi](e^b) = 0. 
\label{eq:homog-infty}
\ee
Next let us evaluate the effect of $e^{b\hat{\Count}}$ on $\hf \phi \cdot h \cdot \phi$. 
Rather than worrying about how to compute the functional derivative of something involving
$\int^{p^2} dq^2 \bigl[1/K(q^2) \bigr]' q^{-2\eta_\star/2}$, we proceed by noticing that~\eq{fp-dual} implies that
\be
	 \hat{D}^+_\star \, \dual_\star[\phi] = \frac{\eta_\star}{2} \phi \cdot \ep^{-1} \cdot \phi,
	 \qquad
	 \Rightarrow
	 \qquad
	  \hat{D}^+_\star \, \hat{\Count} \dual_\star[\phi] = 0.
\ee
This tells us that the vertices of $\hat{\Count} \dual_\star[\phi] $ transform homogeneously with momenta, precisely as those of $\homog[\phi]$ do. Therefore, since $h(p^2)$ contains those---and only those---pieces which do not transform in this way, it must be that
\be
	\hat{\Count}\, \phi \cdot h \cdot \phi = 0,
\ee
from which it follows that
\be
	e^{b\hat{\Count}}
	\hf \phi \cdot h \cdot \phi 
	=
	\hf \phi \cdot h \cdot \phi.
\label{eq:special-h}
\ee
Recalling from~\eq{homog} that $\dual_\star = \homog + \hf \phi \cdot h \cdot \phi$, we obtain, upon employing~\eqs{h-recast}{c}, the crucial result
\be
	e^{b\hat{\Count}}
	\dual_\star[\phi]
	-\hf \phi \cdot \rho \cdot \phi
	=
	\left\{
	\begin{array}{ll}
	\ds	
		e^{b\hat{\Count}} \homog[\phi] , & \eta_\star <2, \ \neq 0,
	\\[1ex]
	\ds
		e^{b\hat{\Count}} \homog[\phi]  - \hf \int_p \phi(p) p^2 \phi(-p), \ & \eta_\star = 0.
	\end{array}
	\right.
\label{eq:crux}
\ee
Finally, by using the results~\eqs{homog-operated}{homog-infty} we find that
\be
	\lim_{b \rightarrow -\infty}
	\Bigl(
	e^{b\hat{\Count}}
	\dual_\star[\phi]
	-\hf \phi \cdot \rho \cdot \phi
	\Bigr)
	= 
	\left\{
	\begin{array}{cl}
	\ds	
		0, & \eta_\star <2, \ \neq 0
	\\ 
	\ds
		-\hf \int_p \phi(p) p^2 \phi(-p), \ & \eta_\star = 0.
	\end{array}
	\right.
\label{eq:crucial}
\ee
As a corollary of this, note that for $\eta_\star=0$, $\lim_{b \rightarrow -\infty} e^{b\hat{\Count}} \dual_\star[\phi] = 0$.

\section{The Correlation Functions}
\label{sec:CFs}

\subsection{Evaluation using the ERG}
\label{sec:Evaluation}

To prove our statement that the only fixed-point with $\eta_\star = 0$ is the Gaussian one, we must understand the relationship between the dual action and the correlation functions, for which we will follow the approach of~\cite{Fundamentals}. The objects that we wish to compute are obtained, as usual, by introducing a source term, $j \cdot \varphi$, at the bare scale \emph{working in the original, dimensionful variables}.%
\footnote{We will deal with the problems associated with IR divergences in $\D=2$ when we encounter them.
}
 Thus
\be
	\pf[j] \sim
	\Fint{\varphi}
	e^{-\Stot_{\Lambda_0}[\varphi] - j\cdot\varphi},
\ee
where the presence of $\sim$ indicates that, if we are either sitting at a critical fixed-point or on a renormalized trajectory, $\Lambda_0$ should be sent to infinity. As usual, the connected correlation functions are computed according to
\be
	G(p_1,\ldots,p_n) \, \deltahat{p_1+\cdots+p_n}
	\sim
	(-1)^n \left. \fder{}{j(p_1)} \cdots \fder{}{j(p_n)} \ln \pf[j] \right\vert_{j=0}.
\label{eq:ConnCorr}
\ee
At the two-point level, we use the shorthand $G(p) \equiv G(p,-p)$.

To calculate these objects using the ERG, we follow the usual strategy of integrating out degrees of freedom between the bare and effective scales. The only difference now is that the effective action develops dependence on $j$. 
Indeed, the flow equation is exactly the same as before, so long as we make the replacement
\be
	\Sint_\Lambda[\varphi] \rightarrow \Tact_\Lambda[\varphi,j] = \Sint_\Lambda[\varphi] + \coupled_{\Lambda}[\varphi,j],
\label{eq:shift}
\ee
where we make the obvious split between the functionals $\Sint$ and $\coupled$, so that
all terms which are independent of $j$ reside in the former. In order that we are computing the correlation functions~\eq{ConnCorr}, we must supply the boundary condition
\be
	\lim_{\Lambda \rightarrow \Lambda_0} \coupled_{\Lambda}[\varphi,j]
	\sim j \cdot  \varphi.
\label{eq:source_bc}
\ee
Substituting~\eq{shift} into~\eq{ConnCorr}, we integrate out all degrees of freedom down to $\Lambda =0$ (at which point the functional integral has been performed). The $\Stot_{\Lambda=0}$ term does not feature
after differentiation \wrt\ the source. This is just as well since $\Stot_{\Lambda=0}$ is divergent,
due to the inverse cutoff function appearing in the two-point vertex.  Since all modes of the field have been integrated over, all field-dependent contributions to $\coupled$ must either vanish or diverge.
We assume that it is the former which is true. Therefore we can write:
\be
	G(p_1,\ldots,p_n) \, \deltahat{p_1+\cdots+p_n}
	\sim
	(-1)^{n+1} \left. \fder{}{j(p_1)} \cdots \fder{}{j(p_n)} \coupled_{\Lambda=0}[0,j] \right\vert_{j=0}.
\label{eq:corr-O_0}
\ee

Thus, the correlation functions are determined by $\coupled_{\Lambda=0}[0,j] $, which can be computed using the flow equation. With this in mind, let us transfer to dimensionless variables, momentarily forgetting that we redefine the field along the flow using $\sqrt{Z}$. For $\varphi$ we recall that this is achieved by writing $\phi(\tilde{p}) = \varphi(p)\Lambda^{(\D+2)/2}$, where we have temporarily reinstated the tilde which explicitly indicates dimensionless momenta. Thus, to ensure that the source term $j \cdot \varphi$ contains no explicit dependence on $\Lambda$, we might conclude that we should rescale $j$ according to $J(\tilde{p}) = j(p) \Lambda^{(\D-2)/2}$. This is partially correct. However, we must not forget that, for each RG step, we additionally redefine the field so as to remove the field strength renormalization from the action. With this in mind, we should take $J(\tilde{p}) = j(p) \Lambda^{(\D-2)/2}\sqrt{Z}$ to ensure that the term $J \cdot \phi$ never has any explicit dependence on the scale.
The flow equation~\eq{Ball} thus becomes:
\be
	\left(
		\partial_t -\hat{D}^- - \hat{D}^J
	\right) \Tact
	=
	\classical{\Tact}{\cutoff'}{\Tact} - \quantum{\cutoff'}{\Tact} 
	- \frac{\eta}{2} \phi \cdot \ep^{-1} \cdot \phi,
\label{eq:Ball-source}
\ee
where (once again dropping the tildes)
\be
	\hat{D}^J = 
	 \int_p
	\biggl[
		\biggl( \frac{\D- 2 + \eta}{2} + p \cdot \partial_p \biggr) J(p)
	\biggr]
	\fder{}{J(p)}.
\label{eq:DJ}
\ee

Our aim now is to consider connected correlation functions at fixed-points. Therefore, we want to find a solution $\partial_t \Tact_\star[\phi,J] = 0$ which, in dimensionful variables, satisfies the boundary condition~\eq{source_bc}. In the first two versions of~\cite{Fundamentals}, the dual action was employed to show that such a solution exists for $\eta_\star <2$, and is given by
\be
	\Tact_\star[\phi,J] = 
	\Sint_\star[\phi]
	+
	\Bigl[
		e^{\bar{J} \cdot (\ep \rho -1 )  \cdot \delta/\delta \phi}
		-1
	\Bigr]
	\Bigl[
	\Sint_\star[\phi] + \hf \phi \cdot \bigl(\ep \rho -1 \bigr)^{-1} \rho \cdot \phi
	\Bigr],
\label{eq:T-solution}
\ee
where $\bar{J}(p) \equiv J(p)/p^2$.
In \app{Alternative} we will demonstrate, for the first time by direct substitution, that this is indeed a solution of~\eq{Ball-source}. Checking that our solution satisfies the boundary condition is easier. First of all, we observe from~\eq{rho-Taylor} that (still in dimensionless variables)
\be
	\frac{\rho(p^2)}{p^2} = 1 + \order{p^2}.
\ee
Now let us transfer to dimensionful variables. For the momenta, we shift $p^2 \mapsto p^2/\Lambda^2$. 
When dealing with the field (recalling the discussion at the end of \sect{flow}) we must remember to undo the effect of $\psi$ along the flow---equivalently, we rescale using the full scaling dimension of the field. Noting that the $\sqrt{Z} \sim (\Lambda/\mu)^{\eta_\star/2}$ coming from this procedure is cancelled by its inverse coming from the source, we find that
\be
	\lim_{\Lambda \rightarrow \infty} \bar{j} \cdot  \rho \cdot \varphi = j \cdot \varphi.
\ee
Moreover, under the transfer to dimensionful variables we have
\begin{align*}
	\bar{j} \cdot \rho \bigl(\ep \rho -1 \bigr)  \cdot \bar{j}
	& =
	\frac{\Lambda^{-2+\eta_\star}}{\mu^{\eta_\star}} \int_p j(p) j(-p) \,
	\frac{\Lambda^2}{p^2} \,
	\order{p^2/\Lambda^2},
\\
	\bar{j} \cdot \bigl(\ep \rho -1 \bigr)  \cdot \fder{}{\varphi}
	& =
	\frac{\Lambda^{-2+\eta_\star}}{\mu^{\eta_\star}} \int_p j(p) \fder{}{\varphi(p)}
	\,  \frac{\Lambda^2}{p^2} 	\order{p^2/\Lambda^2}.
\end{align*}
Since we are working with $\eta_\star <2$, it is apparent that both of these contributions vanish in the limit $\Lambda \rightarrow \infty$, confirming that our solution~\eq{T-solution} satisfies the boundary condition.

There is, however, a problem with the above construction which manifests itself when $\D\leq 2$ and $\eta_\star =0$. This can be amply illustrated by considering the simplest representative of the Gaussian fixed-point, $\Sint_\star =0$. Recalling that, when $\eta_\star = 0$, $\rho(p^2) = p^2$, we have, after translating~\eq{T-solution} into dimensionful variables,
\be
	\coupled_\Lambda[0,j] = \hf \int_p j(p) j(-p) \frac{\cutoff(p^2/\Lambda^2)-1}{p^2}.
\ee
To extract the correlation functions, we must take $\Lambda \rightarrow 0$. Unfortunately, for $\D\leq 2$, this causes the momentum integral to blow up for small $p^2$. The solution to this problem is to compute correlation functions of $\partial_\mu \varphi(x)$, rather than $\varphi(x)$ (just as Pohlmeyer did~\cite{Pohlmeyer}). In this case, we replace~\eq{source_bc} with
\be
	\lim_{\Lambda \rightarrow \Lambda_0} \coupled_{\Lambda}[\varphi,j]
	\sim 
	-i \int_p p_\mu \, j_\mu(p) \varphi(-p).
\ee
Now we repeat the steps leading to~\eq{T-solution} with the appropriate modifications. In particular, we take $J_\mu(p) = j_\mu(p) \Lambda^{\D/2}$, which leads to the factor $\D-2+\eta$ in~\eq{DJ} being replaced by $\D+\eta$. Following \app{Alternative}, it is easy to show that the only modification we need make to~\eq{T-solution} is to replace $\bar{J}(p) = J(p)/p^2$ with $-i J_\mu(p) p_\mu/p^2$.

\subsection{Linearization}

To proceed, we mimic what we did in the sourceless case, and linearize the flow equation. 
For what follows, we will take the source to be a scalar, $J(p)$, which couples to $\phi(p)$ at the bare scale. The translation to the case where the source is a vector, $J_\mu(p)$, which couples to $-ip_\mu \phi(p)$ at the bare scale is obvious and will not be considered separately.
With this in mind, we generalize~\eq{Dual_k}:
\be
	\dualshift_{t,s}[\phi,J]
	\equiv
	- \ln e^{\op_{s-t}} e^{-\Tact_t[\phi,J]},
\label{eq:DualShift_k}
\ee
so that the linearized flow equation (with explicit IR regularization) becomes:
\be
	\biggl(
		\partial_t - \hat{D}^+  - \hat{D}^J  - \frac{\eta}{2} \phi \cdot \ep^{-1} \cdot \phi
		+ \eta \phi \cdot \frac{\cutoff_{s-t}}{\cutoff} \cdot \fder{}{\phi}
		+\frac{\eta}{2} \classical{}{\frac{\cutoff_{s-t} \ep_{s-t}}{\cutoff}}{}
	\biggr) e^{-\dualshift_{t,s}[\phi,J]} = 0.
\label{eq:RescaledDualShiftFlow_k}
\ee
As in the sourceless case, we assume that $\lim_{s \rightarrow \infty} \dualshift_{t,s}[\phi,J]$ exists and understand $\dualshift_t[\phi,J]$ to be given by solutions of 
\be
	\left(
		\partial_t - \hat{D}^+ - \hat{D}^J -\frac{\eta}{2} \phi \cdot \ep^{-1} \cdot \phi
	\right) 
	e^{-\dualshift_t[\phi,J]}
	= 0.
\label{eq:RescaledDualShiftFlow}
\ee
Similarly to before, in the case that $\eta=0$, $\dualshift_{t,\infty}[\phi,J]$ and $\dualshift_t[\phi,J]$ satisfy the same equation and so we identify them. For $\eta \neq 0$, we again tacitly work under the assumption that the same identification can be made.

Now, if we suppose that we take dimensionful $j$ and dimensionful momenta, then it is apparent that (so long as $\phi \cdot \delta \dualshift[\phi,j] /\delta \phi \, \vert_{\phi=0}=0$, which we assume)
\be
	-\flow \lim_{k\rightarrow0} \dualshift_{\Lambda,k}[0,j] = 0,
\ee
where we have traded dependence on $t$ and $s$ for $\Lambda$ and $k$.
So let us return to~\eq{DualShift_k}, but  with the source and all momenta rendered dimensionful and the field set to zero. We have just learnt that $ \lim_{k\rightarrow0} \dualshift_{\Lambda,k}[0,j]$ is independent of scale, so let us evaluate it at $\Lambda=0$. With dimensionful momenta, 
\[
\op_{\Lambda,k} \equiv \int_p \fder{}{\phi(p)} \frac{\cutoff(p^2/\Lambda^2) - \cutoff(p^2/k^2)}{p^2} \fder{}{\phi(-p)}.
\]
 But by saying that we can identify $\lim_{k\rightarrow0} \dualshift_k[\phi,j]$ with solutions of~\eq{RescaledDualShiftFlow} we are effectively saying that  we can remove the IR regularization from $\dualshift_k[\phi,J]$ without picking up any corrections that survive the $k \rightarrow 0$ limit. But if this is true, then it implies that there are not any IR effects which spoil subsequently taking the limit $\Lambda \rightarrow 0$. Since $\lim_{\Lambda \rightarrow 0} \cutoff(p^2/\Lambda^2) = 0$ we therefore conclude that
\be
	\lim_{k\rightarrow0} \dualshift_{\Lambda,k}[0,j]  = \dualshift[0,j] 
	= \lim_{\Lambda \rightarrow 0} \dualshift[0,j] = \Tact_{\Lambda=0}[0,j]
	=
	\coupled_{\Lambda=0}[0,j],
\label{eq:equal}
\ee
where, in the last step, we have used the decomposition~\eq{shift} and thrown away the
constant coming from $\Sint_{\Lambda=0}[0]$.%
\footnote{For $\eta_\star =0$, there is a clearer way to proceed, for here we find that $-\flow \dualshift_k[0,j] = 0$, without the need to send $k \rightarrow 0$. Evaluating $\dualshift_k[0,j]$ at $\Lambda = k$ we see from~\eq{DualShift_k} that $\dualshift_k[0,j] = \Tact_k[0,j]$. Now sending $k \rightarrow 0$, we recover the conclusion of~\eq{equal}.}
Combining this with~\eq{corr-O_0},  the correlation functions can be written as
\be
	G(p_1,\ldots,p_n) \, \deltahat{p_1+\cdots+p_n}
	\sim
	\left.
	(-1)^{n+1} \fder{}{j(p_1)} \cdots \fder{}{j(p_n)} 
	\dualshift[0,j]
	\right\vert_{j=0}.
\label{eq:correlations-dual}
\ee

With this in mind, let us return to dimensionless variables and derive an expression for $\dualshift_\star[\phi,J]$ using our expression for $\Tact_\star[\phi,J]$ given by \eq{T-solution}. To err on the side of caution, let us first compute $\dualshift_{t,s}[\phi,J]$:
\be
	\dualshift_{t,s}[\phi,J] = 
	\hf \bar{J} \cdot \rho \bigl(\ep \rho -1 \bigr)  \cdot \bar{J}
	-\ln
	\Bigl\{
		e^{\op_{s-t}}
		\exp
		\Bigl(
			-\bar{J} \cdot  \rho \cdot \phi
			- e^{\bar{J} \cdot (\ep \rho -1 )  \cdot \delta/\delta \phi} \Sint_\star[\phi]
		\Bigr)
	\Bigr\}.
\ee
This can be processed by recognizing that
\begin{align*}
	e^{\op_{s-t}} e^{-\bar{J} \cdot  \rho \cdot \phi}
	& =
	\exp\Bigl(
	-\bar{J} \cdot  \rho \cdot \phi + \hf \bar{J} \cdot  \rho^2 \ep_{s-t} \cdot \bar{J} 
	- \bar{J} \cdot \rho \ep_{s-t} \cdot \delta /\delta \phi
	\Bigr)
	e^{\op_{s-t}},
\\
	\exp 
	\Bigl(
		- e^{\bar{J} \cdot (\ep \rho -1 )  \cdot \delta/\delta \phi} \Sint_\star[\phi]
	\Bigr)
	& =
	e^{\bar{J} \cdot (\ep \rho -1 )  \cdot \delta/\delta \phi}  e^{-\Sint_\star[\phi]},
\end{align*}
upon which it is apparent that
\be
	\dualshift_{t,s}[\phi,J] = 
	\bar{J} \cdot  \rho \cdot \phi 
	-\hf \bar{J} \cdot \rho \bigl[1 + \rho \bigl(\ep_{s-t} - \ep \bigr) \bigr] \cdot \bar{J} 
	-\ln 
	e^{-\bar{J} \cdot [1 + \rho (\ep_{s-t} - \ep ) ] \cdot \delta/\delta \phi} e^{-\dual_{t,s}[\phi]}.
\ee
Therefore,
\be
	\dualshift_{t,s}[\phi,J]  - \dual_{t,s}[\phi]
	=
	\Bigl\{
		e^{-\bar{J} \cdot [1 + \rho (\ep_{s-t} - \ep ) ] \cdot \delta/\delta \phi} - 1
	\Bigr\}
	\biggl\{
		\dual_{t,s}[\phi] - \hf \phi \cdot \frac{\rho}{1 + \rho (\ep_{s-t} - \ep ) } \cdot \phi
	\biggr\}.
\ee
Taking the limit $s\rightarrow \infty$, and recalling that $\ep_{\infty}(p^2) = \ep(p^2)$ it is clear---given the usual assumption(s) concerning the existence and form of $\lim_{s\rightarrow\infty}\dual_{s,t}[\phi]$---that
\be
	\dualshift_\star[\phi,J]  - \dual_\star[\phi]
	=
	\Bigl(
		e^{-\bar{J} \cdot \delta/\delta \phi} - 1
	\Bigr)
	\Bigl(
		\dual_\star[\phi] - \hf \phi \cdot \rho\cdot \phi
	\Bigr).
\ee

Finally, recalling from~\eq{correlations-dual} that the correlation functions are evaluated at vanishing field it is apparent that (in dimensionless variables) 
\begin{multline}
	G(p_1,\ldots, p_n) \, \deltahat{p_1 + \cdots + p_n}
	=
\\
	\left.
	(-1)^{n+1} \fder{}{J(p_1)} \cdots \fder{}{J(p_n)} 
	\Bigl(
		e^{-\bar{J} \cdot \delta/\delta \phi} - 1
	\Bigr)
	\Bigl(
		\dual_\star[\phi] - \hf \phi \cdot \rho \cdot \phi
	\Bigr)
	\right\vert_{J,\phi=0}.
\label{eq:CorrFnsByDual}
\end{multline}
(Since we are at a fixed-point, the bare scale can and has been sent to infinity, allowing us to replace the usual $\sim$ with an $=$.)
Let us note that for $\eta_\star <2, \ \neq 0$, it is straightforward to check that positivity of the two-point  connected correlation function requires $\intconst_{\eta_\star} >0$, which is the reason for the choice of sign in~\eq{constants}. It is also well worth pointing out that the structure of $\dual_\star$ is such that the connected correlation functions transform covariantly under dilatations~\cite{Fundamentals}. This is a strong indication that the assumptions we have made concerning the existence and form of $\lim_{s\rightarrow \infty} \dual_{t,s}[\phi]$ are justified.

\section{Extending Pohlmeyer's Theorem}
\label{sec:Pohl}

Given a fixed-point action, $\Sint_\star[\phi]$, \eqn{CorrFnsByDual} provides a recipe for computing the connected correlation functions in terms of the associated dual action. However, from \eq{line} we know that there exists a line of fixed-points, $e^{b \hat{\Count}} \Sint_\star[\phi]$, for which the dual action is given by~\eq{line-dual}. Therefore, the expression for the connected correlation functions along the line is:
\begin{multline}
	G(p_1,\ldots, p_n;b) \, \deltahat{p_1 + \cdots + p_n}
	=
\\
	\left.
	(-1)^{n+1} \fder{}{J(p_1)} \cdots \fder{}{J(p_n)} 
	\Bigl(
		e^{-\bar{J} \cdot \delta/\delta \phi} - 1
	\Bigr)
	\Bigl(
		e^{b \hat{\Count}} \dual_\star[\phi] - \hf \phi \cdot \rho \cdot \phi
	\Bigr)
	\right\vert_{J,\phi=0}.
\label{eq:CorrFnsLine}
\end{multline}
Let us start by taking $\eta_\star < 2, \ \neq 0$. It is obvious from~\eqs{crux}{homog-operated} that,
for finite values of $b$, running along the line of fixed-points corresponds to changing the normalization of the field. This is precisely what we expect since, for  $\eta_\star < 2, \ \neq 0$, our exactly marginal operator is redundant~\cite{Fundamentals}. Now consider sending $b \rightarrow -\infty$. Using~\eq{crucial}, it is apparent that, for 
$\eta_\star < 2, \ \neq 0$, \emph{all} connected correlation functions vanish in this limit. This is not acceptable, and so we conclude that $-\infty < b <\infty$, in this case. 

However, things are very different when $\eta_\star = 0$. First of all, the extra term in~\eq{crux}---as compared with the $\eta_\star <2, \ \neq 0$ case---means that running along the line of fixed-points does \emph{not} amount to changing the normalization of the field (unless $\homog$ has only a two-point contribution). Recall that, in this case, $\rho(p^2) = p^2$ and the two-point contribution to the dual action is just 
$-\intconst_0 e^{b}\, \hf \int_p \phi(p) p^2 \phi(-p)$. Presuming that $\intconst_0 >-1$, which ensures that $G(p;0)$ is positive, it is apparent 
that the  two-point connected correlation function survives the limit (note that we are taking $\D>2$):
\be
	\lim_{b\rightarrow -\infty} G(p;b)= \frac{1}{p^2}, \qquad \eta_\star =0.
\ee
This is perfectly acceptable: it simply tells us that as we move along the line of fixed-points generated by 
$e^{b \hat{\Count}}$, we ultimately sink into a theory with the same correlation functions as Gaussian one as $b \rightarrow -\infty$. Moreover, it is easy to see that this theory not only has the same correlation functions as the Gaussian theory but must \emph{be} the Gaussian theory. For as noted under~\eq{crucial}, when we take the limit $b \rightarrow -\infty$ for a theory with $\eta_\star = 0$, we find that $\dual_\star[\phi]$ vanishes. However, from the definition of the dual action, \eq{Dual}, it is immediately apparent that this corresponds to the Gaussian fixed-point.

But now we arrive at a contradiction, for we know that at the Gaussian fixed-point $e^{b \hat{\Count}}$ simply generates motion along the exactly marginal, redundant direction. Indeed, the connected correlation functions along the associated line are given by:
\begin{align}
	G_{\mathrm{Gaussian}}(p;b) & = (1+ \intconst_0 e^b) \frac{1}{p^2},
\\
	G_{\mathrm{Gaussian}}(p_1,\ldots,p_n;b) & = 0, \qquad n>2.
\end{align}
[Recalling~\eqs{Gaussian}{repeated}, let us note that $B_0 = -B$, as can be confirmed by using~\eq{T-solution}, together with~\eqs{shift}{corr-O_0}].
Obviously, the $n>2$-point  connected correlation functions vanish not just for $b \rightarrow -\infty$, but for all values of $b$: there is no notion of these higher-point connected correlation functions sinking into their Gaussian values of zero only as $b \rightarrow -\infty$. Therefore, we conclude that the only resolution to our contradiction is that any fixed-point with $\eta_\star = 0$ is the Gaussian one. This result is true in general dimension.

Note that were we to relax the positivity constraint, $\intconst_0 > -1$, the above argument would break down: for then there would be some value of $b$ with $0>b >-\infty$ for which the two-point connected correlation function vanishes, impeding the limit $b \rightarrow -\infty$.

Finally, for $\D \leq 2$, we instead consider correlation functions of $\partial_\mu \phi(x)$. The effects on~\eq{CorrFnsLine} are that the $\delta/\delta J(p_i)$ are replaced by $\delta/\delta J_{\mu_i}(p_i)$
and $\bar{J}(p)$ is interpreted as $-i J_\mu(p) p_\mu/p^2$. The conclusion that the only fixed-point with $\eta_\star =0$ is the Gaussian one remains the same.

\section{Conclusion}
\label{sec:Conclusion}

Having completed the extension of Pohlmeyer's theorem, let us unpick what we have done. The steps of the argument (glossing over the subtleties associated with IR divergences) are as follows:
\begin{enumerate}
	\item \Eqn{line} recalls that every fixed-point, $\Sint_\star[\phi]$, 
	is a member of a family generated by the action of 
	$e^{b\hat{\Count}}$. Defining the dual action 
	$\dual[\phi] \equiv -\ln\bigl\{ \exp \bigl(\hf \fder{}{\phi} \! \cdot \!{\scriptstyle \ep} \! \cdot \!  \fder{}
	{\phi} \bigr) e^{-\Sint[\phi]} \bigr\}$,
	the act of running along this line induces a change $e^{b\hat{\Count}} \dual_\star[\phi]$, as noted 
	in~\eq{line-dual}.
	
	\item Correlation functions at a critical fixed-point (requiring $\eta_\star <2$) follow from the 
	solution~\eq{T-solution} to the source-dependent flow equation. This solution satisfies the requisite
	boundary condition that \emph{in dimensionful variables} the source-dependent part of the action 
	reduces to $j \cdot \varphi$ in the limit $\Lambda \rightarrow \infty$.

	\item Using~\eq{T-solution}, the correlation functions at a critical fixed-point are written in terms of 
	the dual action, $\dual_\star[\phi]$, in~\eq{CorrFnsByDual}.
	
	\item Running along the line of fixed-points generated by $e^{b \hat{\Count}}$ it is observed, 
	from~\eqs{crux}{crucial}, that:
	\begin{enumerate}
		\item For $\eta_\star \neq 0$, $b > -\infty$, else all correlation functions vanish;
		
		\item For $\eta_\star = 0$, the limit $b \rightarrow -\infty$ 
		(which can only be taken if the two-point connected correlation function is positive for $b=0$) 
		causes the correlation functions
		to sink into those of the Gaussian fixed-point. This in turn implies that the action itself
		becomes the Gaussian action.
		\label{killer}
	\end{enumerate}
	
	\item However, acting on the Gaussian fixed-point, $e^{b \hat{\Count}}$ generates a line of 
	\emph{equivalent} fixed-points, for which the correlation functions are precisely the Gaussian ones 
	(up to a multiplicative factor), even for $b > -\infty$. The only way to reconcile~\ref{killer} with
	this fact is if the only fixed-point with $\eta_\star = 0$, and for which $G(p;0)>0$,
	is the Gaussian one.
\end{enumerate}

Although the analysis of this paper has been carried out for theories of a single scalar field, it should be apparent that what we have done can be easily generalized to theories containing an arbitrary selection of non-gauge fields (though application to non-linear sigma models would necessitate developing the methodology). Essentially, all we need to do is reinterpret $\phi$ as appropriate, inserting indices wherever necessary. Indeed, if the `dot-notation' employed in the flow equation~\eq{Ball} is taken to include a summation over any indices present, then we are basically done. (Supersymmetric theories can readily be treated using the superfield formalism, should one so desire~\cite{Susy-Chiral}.)

Gauge fields, however, present their own difficulties. Indeed, reconciling gauge invariance with the cutoff inherent in the ERG is a non-trivial problem. The conventional approach is to allow the cutoff to break gauge invariance (in one way or another), but in such a way that it is recovered in the limit $\Lambda \rightarrow 0$ \ie\ the limit in which all quantum fluctuations have been integrated out (see~\cite{JMP-Review} for a comprehensive review). Such an approach looks unlikely to blend well with the methods used in this paper. There is, however, a radical alternative pioneered by Morris~\cite{ym1,aprop}.

Using a version of covariant higher derivative regularization, it is in fact possible to construct a gauge invariant cutoff appropriate for use within the ERG~\cite{sunn}. More remarkably, it turns out that there exist ERGs for which gauge invariance  is manifest: at no point is any gauge fixing performed. Such a scheme has been constructed for QED~\cite{qed}, $\SU(N)$ Yang-Mills~\cite{mgierg1,mgierg2}  and QCD~\cite{qcd}. The price to pay for manifest gauge invariance is considerable complication of the flow equation, which in itself will necessitate developing the techniques employed in this paper if progress is to be made. Not only this, but the standard correlation functions---which play an essential role in the above analysis---do not appear in the formalism since they are not manifestly gauge invariant: only expectation values of gauge invariant operators are appropriate things to compute~\cite{ym1,evalues,univ}.

Whilst the challenges of adapting the methods of this paper to gauge theory are thus considerable, the manifestly gauge invariant approach has a tantalizing feature: in both the non-Abelian case and the Abelian case with matter, the anomalous dimension of the gauge field is automatically 
zero~\cite{ym1,qed}.  Therefore, if we suppose for a moment that the results of this paper extend to gauge theories, then this would immediately tell us that the only renormalizable gauge theories are those that can be constructed about the Gaussian fixed-point. Since this fixed-point does not have any relevant or marginally relevant directions for $\D >4$, this would imply that there is no possibility of a field-theoretic UV completion of higher-dimension gauge theories. Investigating this further is clearly worthwhile.

\begin{acknowledgments}
	I would like to thank Joe Polchinski for bringing to my attention the importance of positivity and 
	Hugh Osborn for comments on the manuscript. 
	Thanks also to Sebastian J\"{a}ger for helpful discussions on $\D$-dimensional integrals.
	This work was supported by the Science and Technology Facilities Council 
	[grant number ST/F008848/1].
\end{acknowledgments}

\appendix

\section{The Solution for $\Tact_\star$}
\label{app:Alternative}

In this appendix we will prove that~\eq{T-solution} solves the source-dependent flow equation at a fixed-point. As a warmup, we will first prove~\eq{line}. Recalling~\eqs{fp-operator}{classifier} let us now consider
\be
	\fpop(e^{b\hat{\Count}} \Sint_\star) 
	=
	\classifier 
	\bigl(e^{b\hat{\Count}} -1 \bigr) \Sint_\star
	+
	\classical{\bigl(e^{b\hat{\Count}} -1 \bigr)\Sint_\star}{\cutoff'}{\bigl(e^{b\hat{\Count}} -1 \bigr)\Sint_\star},
\label{eq:zero}
\ee
where we have used~\eq{fp-condition}. We would like to show that this expression vanishes. The simplest way to do this is to start by differentiating \wrt\ $b$:
\be
	\der{}{b}\, \fpop(e^{b\hat{\Count}} \Sint_\star)	=
	\classifier \, e^{b\hat{\Count}} \hat{\Count} \Sint_\star
	+
	2
	\classical{\bigl(e^{b\hat{\Count}} -1 \bigr)\Sint_\star}{\cutoff'}{\, e^{b\hat{\Count}} \hat{\Count} \Sint_\star}.
\label{eq:differentiated}
\ee
Using~\eq{emop}, which tells us that $\classifier \, \hat{\Count} \Sint_\star = 0$, we see that
\be
	\classifier \, e^{b\hat{\Count}} \hat{\Count} \Sint_\star
	=
	\bigl[\classifier , e^{b\hat{\Count}}\bigr] \hat{\Count} \Sint_\star.
\ee
The commutator can be processed using standard tricks:
\begin{align}
	\bigl[
		\classifier, e^{b \hat{\Count}}
	\bigr]	
	& =
	\int_0^1 e^{sb  \hat{\Count}}
	\bigl[
		\classifier, b \hat{\Count}
	\bigr]	
	e^{-sb \hat{\Count}}
	e^{b \hat{\Count}}
\nonumber
\\
	& =
	-\Bigl(
	\bigl[
		 b\hat{\Count}, \classifier
	\bigr]	
	+
	\frac{1}{2!} 
	\bigl[
	b\hat{\Count},
	\bigl[
		 b\hat{\Count}, \classifier
	\bigr]
	\bigr]	
	+
	\frac{1}{3!}
	\bigl[ 
	b\hat{\Count},
	\bigl[
	b\hat{\Count},
	\bigl[
		 b\hat{\Count}, \classifier
	\bigr]
	\bigr]	
	\bigr]
	+\ldots
	\Bigr) e^{b \hat{\Count}}
\nonumber
\\
	& =
	-2\classical{\bigl(e^{b\hat{\Count}} -1\bigr)\Sint_\star}{\cutoff'}{} e^{b\hat{\Count}},
\end{align}
where the last line is obtained using
\be
	\bigl[\hat{\Count}, \classifier\bigr]
	=
	2\classical{\hat{\Count} \Sint_\star}{\cutoff'}{}.
\label{eq:commutator}
\ee
It is thus immediately apparent that the \rhs\ of~\eq{differentiated} vanishes. Integrating up, the integration constant can be seen to be zero by noting that the \rhs\ of~\eq{zero} vanishes for $b=0$.
Therefore,
\be
	\fpop(e^{b\hat{\Count}} \Sint_\star) 
	=0,
\label{eq:demonstrated}
\ee
which implies that any fixed-point is a member of a line generated by $e^{b\hat{\Count}}$.
\Eqn{line} follows after demanding that the result of acting with $e^{(b'+b)\hat{\Count}}$
is the same as acting with $e^{b\hat{\Count}}$ followed by $e^{b'\hat{\Count}}$, for some real parameter $b'$.

With this proof in mind, we would now like to consider solutions to the fixed-point equation
\be
	\fpop_J(\Tact_\star[\phi,J]) = 0,
\label{eq:fp-source}
\ee
where
\be
	\fpop_J(\Tact_\star) =
	\classical{\Tact_\star}{\cutoff'}{\Tact_\star} - \quantum{\cutoff'}{\Tact_\star} 
	+ \Bigl(\hat{D}^-_\star +  \hat{D}^J_\star \Bigr) \Tact_\star
	- \frac{\eta_\star}{2} \phi \cdot \ep^{-1} \cdot \phi.
\ee
Introducing some as yet undetermined function $g(p^2)$, let us define
\be
	\hat{\mathcal{R}} \equiv \bar{J} \cdot \bigl(\ep g -1 \bigr) \cdot \delta / \delta \phi,
	\qquad
	\Stilde_\star \equiv \Sint_\star + \hf \phi \cdot \bigl(\ep g -1 \bigr)^{-1} g \cdot \phi,
	\qquad
	\coupled_a[\phi,J]
	\equiv
	\bigl(e^{a \hat{\mathcal{R}}} - 1 \bigr) 
	\Stilde_\star.
\ee
Using the fact that $\fpop_J(\Sint_\star[\phi]) = 0$, we have that
\be
	\fpop_J(\Sint_\star + \coupled_a)
	=
	\classifier_J \coupled_a
	+ \classical{\coupled_a}{\cutoff'}{\coupled_a},
\label{eq:analogue}
\ee
where, recalling~\eq{classifier},
\be
	\classifier_{J} = \classifier + \hat{D}^J_\star =
	2 \classical{\Sint_\star}{\cutoff'}{} - \quantum{\cutoff'}{} + \hat{D}^{-}_\star +  \hat{D}^J_\star.
\ee
Notice that~\eq{analogue} is the analogue of~\eq{zero}. To build on this analogy, let us note that
\be
	\bigl[
		\hat{\mathcal{R}}, \hat{D}^{-}_\star +  \hat{D}^J_\star
	\bigr]
	=
	\int_p \frac{J(p)}{p^2}
	\Bigl[
		p \cdot \partial_p \, \ep(p^2) g(p^2) - \eta_\star \ep(p^2) g(p^2) +\eta_\star
	\Bigr]
	\fder{}{\phi(p)}
\ee
from which it is easy to check that, so long as $g(p^2)$ satisfies
\be
	- \frac{2+\eta_\star}{2} g(p^2) + p^2 g'(p^2) +\frac{\eta_\star}{2} \ep^{-1}(p^2) =0,
\label{eq:g-condition}
\ee
we have (up to a neglected constant in the second case)
\be
	\bigl[
		 \hat{\mathcal{R}}, \classifier_J
	\bigr]	
	=
	2\classical{\, \hat{\mathcal{R}}\Stilde_\star}{\cutoff'}{},
	\qquad
	\classifier_{J} \,
	\hat{\mathcal{R}}
	\Stilde
	= 0.
\ee
These equations are in correspondence with~\eqs{commutator}{emop}, respectively. Now we simply repeat the steps leading to~\eq{demonstrated} (though this time differentiating \wrt\ $a$), from which it follows that
a fixed-point solution to the source-dependent flow equation is
\be
	\Tact_\star[\phi,J] = \Sint_\star[\phi] + 
	\Bigl[
		e^{a \bar{J} \cdot (\ep g -1) \cdot \delta / \delta \phi}
		- 1
	\Bigr]
	\Bigl[
		\Sint_\star + \hf \phi \cdot \bigl(\ep g -1 \bigr)^{-1} g \cdot \phi 
	\Bigr].
\ee
The final step is to ensure that the boundary condition~\eq{source_bc} is satisfied. To this end, we must do two things. First, we set $a=1$. Secondly, we observe by comparing~\eqs{h-eq}{g-condition} that $g(p^2)$ and $h(p^2)$ satisfy the same equation. Therefore, for some constant $\alpha_{\eta_\star}$,
\be
	g(p^2) = h(p^2)  + \alpha_{\eta_\star} p^{2(1+\eta_\star/2)}.
\ee
Looking at~\eq{h-recast} we conclude that a valid choice for $g$ is $g(p^2) = \rho(p^2)$, recovering~\eq{T-solution}.

\bibliography{../../Biblios/ERG.bib,../../Biblios/Misc.bib,../../Biblios/Books.bib,../../Biblios/NC.bib}

\end{document}